\documentclass[twocolumn]{aastex63}

\usepackage[english]{babel}

\usepackage{amsmath}
\usepackage{graphicx}

\shorttitle{Milli-Hz GW background produced by QPEs}
\shortauthors{Chen et al.}

\begin{document}
\title{Milli-Hertz Gravitational Wave Background Produced by Quasi-Periodic Eruptions}
\author[0000-0003-3950-9317]{Xian Chen}
\affiliation{Astronomy Department, School of Physics, Peking University, Beijing 100871, China}
\affiliation{Kavli Institute for Astronomy and Astrophysics, Peking University, Beijing 100871, China}
\email{xian.chen@pku.edu.cn}

\author{Yu Qiu}
\affiliation{Astronomy Department, School of Physics, Peking University, Beijing 100871, China}

\author[0000-0001-6530-0424]{Shuo Li}
\affil{National Astronomical Observatories, Chinese Academy of Sciences, Beijing 100012, China}

\author{F. K. Liu}
\affiliation{Astronomy Department, School of Physics, Peking University, Beijing 100871, China}
\affiliation{Kavli Institute for Astronomy and Astrophysics, Peking University, Beijing 100871, China}

\begin{abstract}
Extreme-mass-ratio inspirals (EMRIs) are important targets for future
space-borne gravitational-wave (GW) detectors, such as the Laser Interferometer
Sapce Antenna (LISA).  Recent works suggest that EMRI may reside in a
population of newly discovered X-ray transients called ``quasi-periodic
eruptions'' (QPEs).  Here we follow this scenario and investigate the
detectability of the five recently discovered QPEs by LISA.  We consider two
specific models in which the QPEs are made of either stellar-mass objects
moving on circular orbits around massive black holes (MBHs) or white dwarfs
(WDs) on eccentric orbits around MBHs.  We find that in either case each QPE is
too weak to be resolvable by LISA. However, if QPEs are made of eccentric
WD-MBH binaries,  they radiate GWs in a wide range of frequencies. The broad
spectra overlap to form a background which, between $0.003-0.02$ Hz, exceeds
the background known to exist due to other types of sources.  Presence of this
GW background in the LISA band could impact the future search for the seed
black holes at high redshift as well as the stellar-mass binary black holes in
the local universe.  
\end{abstract}


\section{Introduction}\label{sec:intro}

An extreme-mass-ratio inspiral (EMRI) consists of a massive black hole (MBH)
and a small compact object, such as a stellar-mass black hole (BH), a neutron
star, or a white dwarf (WD), moving on a tightly bound orbit
\citep{amaro-seoane07}.  Because of gravitational-wave (GW) radiation, the
orbit decays and the small object eventually coalesces with the MBH.  If the
mass of the MBH is $10^5-10^7M_\odot$, the GW radiated during the last few
years of the system falls in the sensitive band of the Laser Interferometer
Space Antenna \citep[LISA,][]{lisa17}. During this period, as many as
$10^4-10^5$ GW cycles could be accumulated in the data stream, providing rich
information about the spacetime geometry close to a MBH \citep{gair13,berry19}.

Despite their scientific importance, many basic properties of EMRIs, such as
the event rate,  are largely unconstrained. The difficulty lies in the lack of
a distinctive electromagnetic (EM) signature.  For example, the EMRIs
containing stellar-mass BHs are considered to dominate the
EMRI population
\citep{pacheco06}, but the predicted event rate varies from one dozen per year
(within a redshift of $z=4.5$) to as high as  a few $\times10^4$ per year
\citep[see][for a summaries]{babak17,gair17}.  The redshift distribution is also
uncertain.  If most EMRIs are at high redshift, they would form a GW
background which is practically indistinguishable from noise
\citep{sigl07,bonetti20}.

Unlike stellar-mass BH or neutron star, a WD revolving around a MBH could be
tidally detonated if the MBH has moderate mass \citep[$10^3-10^6M_\odot$,
e.g.][]{luminet89,rosswog09}, or it could activate the MBH via Roche-lobe
overflow and tidal disruption \citep[e.g.][]{ivanov07,zalamea10,macleod14}.
Therefore, the EMRIs containing WDs are potential targets for joint EM and GW
observations \citep{sesana08}. In particular, they encode valuable information
about the astrophysical environments which lead to the formation of EMRIs.  In
theory, various dynamical processes could deliver WDs to the vicinity of MBHs,
including dynamical relaxation of star cluster \citep{hils95}, tidal capture
\citep{ivanov07}, partial disruption of red giant stars \citep{bogdanovic14},
and tidal separation of WD binaries \citep{miller05}.  It is estimated that as
many as $10^2$ such EMRIs could be detected by LISA with reasonable
signal-to-noise ratio \citep[SNR,][]{hils95,sigurdsson97,ivanov02,sesana08}.

Interestingly, the EM counterpart to the above WD EMRI may have been found in a
new type of transient called ``quasi-period eruption''
\citep[QPE,][]{2019Natur.573..381M,giustini20,2021Natur.592..704A,2021ApJ...921L..40C}.
Five QPEs have been discovered so far and they share a distinctive feature:
within an hour the X-ray count rate surges by one to two orders of magnitude
and such an eruption recurs every few hours.  The short duration of each
outburst and the short recurrence timescale resemble the characteristics of a
small object swooping by a MBH periodically along a tightly bound, highly
eccentric orbit.  The similarity leads to the suggestion that QPEs are powered
by eccentric WD EMRIs whose WDs are filling up their Roche lobes and feeding
the MBHs during their pericenter passages
\citep{2020MNRAS.493L.120K,2021arXiv210903471Z}.  This interpretation is
further supported by the observational evidence of earlier tidal disruption
events in two of the QPEs
\citep{2019Natur.573..381M,sheng21,2021ApJ...921L..40C}, corroborating the
picture that partial disruption of stars could deposit their compat cores (such
as WDs) to the close vicinity of MBHs.

Further theoretical studies suggest that WD-MBH binaries may be too short-lived
to explain the detection rate of QPEs because the WDs would expand in a runaway
fashion as soon as the mass transfer starts \citep{2021arXiv210713015M}.  One
way of alleviating the problem invokes less eccentric orbits for those objects
(not necessarily WDs) around MBHs, so that mass transfer can be avoided
\citep{2021arXiv210713015M,2021ApJ...921L..32X,2021MNRAS.503.1703I}.  In these
models, QPEs also emit GWs because they are essentially still EMRIs.  Another
class of models do not rely on EMRIs but attribute the X-ray eruption to the
instability of accretion disk
\citep{2019Natur.573..381M,motta20,sniegowska20,2021ApJ...909...82R}.  In this
case, little GW radiation is expected. These models, however, have difficulties
explaining two of the QPEs which are found in quiescent galaxies showing no
sign of accretion disks \citep{2021Natur.592..704A}.

Although QPEs may contain EMRIs, whether they can be detected by LISA is still unclear. 
The conventional way of evaluating the
detectability of a GW source by its characteristic strain 
\citep[see, e.g.,][for WD EMRIs]{sesana08,2021arXiv210903471Z} could be insufficient for QPEs.  First, the
characteristic strain is useful, i.e., it is positively correlated with the SNR, only if the GW
frequency increases rapidly during the observational period. 
Such a signal is called ``chirp signal''
\citep[e.g.][]{robson19}.
This is not the case for QPEs because, if they are EMRIs, their orbits evolve on a timescale of
$10^3-10^4$ years according to the previous studies
\citep[e.g.][]{2020MNRAS.493L.120K,2021arXiv210713015M,2021MNRAS.503.1703I}.
Such a timescale is much longer than the mission duration of LISA. As a result, the SNR is
suppressed substantially.  Second, when the orbital eccentricity is high, as
would be the case if QPEs are powered by WDs
\citep[e.g.][]{2020MNRAS.493L.120K}, the GW power is emitted in a wide range of
harmonic frequencies \citep{peters63}. Only those harmonics falling in the
sensitive band of LISA contribute to the SNR.  Third, recent study of the EMRIs
with stellar BHs suggests that although the majority are unresolvable by LISA,
together they form a background which may be higher than the instrument noise
\citep{bonetti20}.  Whether QPEs produce a similar background deserves
investigation. Understanding this background is important because it
may impinge on the detection of the seed MBHs at high redshift, as well as the
binary BHs (BBHs) which could be the progenitors of the sources already
detected by the Laser Interferometer Gravitational-wave Observatory (LIGO) and
the Virgo detectors \citep[e.g.][]{bonetti20}.

Here we take the above three factors into account and study the detectability of
QPEs by the future LISA mission. The paper is organized as follows.
In Section~\ref{sec:models}, we describe two models proposed
for QPEs which contain EMRIs. Based on these models,  we calculate the corresponding 
GW spectra for the five detected QPEs.
In Section~\ref{sec:GWbg} we compute the GW background formed by QPEs and
investigate its detectability by LISA. We also compare it 
with the GW background due to other types of sources and 
evaluate the impact on the
future search of seed BHs and BBHs by LISA.
In Section~\ref{sec:dis} summarize our results and discuss the caveats.  
Throughout the paper, we assume a standard $\Lambda$CDM cosmology with the parameters
$H_0=70\,{\rm km\,s^{-1}\,Mpc^{-1}}$, $\Omega_\Lambda=0.7$, and $\Omega_M=0.3$.

\section{Models}\label{sec:models}

\subsection{EMRIs on circular orbits}

We first consider a model in which QPEs are produced by stellar-mass objects
moving on relatively circular orbits around MBHs
\citep{2021arXiv210713015M,2021ApJ...921L..32X}. Such an orbit has three
parameters, the mass of the MBH $M$, the mass of the stellar-mass object $m$,
and the orbital period $P$.  For the five QPEs detected so far, we give their
parameters in Table~\ref{tab:sample} which are derived in the following ways.

The orbital period $P$ is determined by the time interval between successive
eruptions.  Note that in some models, the small object collides with the
accretion disk of the MBH twice per orbital period, and hence $P$ is twice the
time interval between eruptions \citep{2021ApJ...921L..32X}. We neglect this
factor of two because it does not qualitatively affect the amplitude and
detectability of the GWs.

\begin{table*}
\centering
\caption{\label{tab:sample}QPE sample and their parameters}
\begin{tabular}{ccccccccc}
\hline
 Source & $z$ & $M/M_{\odot}$ & $P/\mathrm{ks}$ & $\Delta t/\mathrm{ks}$ & $L/\mathrm{erg}\ \mathrm{s}^{-1}$ & $m/M_\odot$ & $e$ &Ref. \\
\hline
 GSN 069 & 0.018 & $4.0\times 10^5$ & $31.55$ & 2.05 & $5.0\times10^{42}$ & 0.322&0.972 & \cite{2019Natur.573..381M} \\
RX J1301.9+2747 & 0.02358 & $1.8\times10^6$ & $16.5$ & 1.2 & $1.4\times10^{42}$ &0.150&0.928& \cite{giustini20} \\
eRO-QPE1 & 0.0505 & $9.1\times10^5$ & 66.6 & 13.7 & $3.3\times10^{42}$ &0.461&0.986& \cite{2021Natur.592..704A} \\
eRO-QPE2 & 0.0175 & $2.3\times10^5$ & 8.64 & 0.8 & $1.0\times10^{42}$ &0.178&0.901& \cite{2021Natur.592..704A} \\
XMMSL1 J024916.6-041244 & 0.019 & $8.5\times10^4$ & 9 & 1 & $3.4\times10^{41}$ &0.169&0.901& \cite{2021ApJ...921L..40C}\\
\hline
\end{tabular}
\end{table*}

The mass $M$ of the MBH is not an observable and is derived for different QPEs
using different methods.  For GSN 069 and RX J1301.9+2747, the masses are
derived from fitting their X-ray spectra with accretion disk models
\citep[adopted from][]{2019Natur.573..381M,2017ApJ...837....3S}.  The mass of
XMMSL1 J0249-041244 is inferred from the correlation between the mass of a MBH
and the velocity dispersion of the bulge of the host galaxy
\citep[from][]{2019MNRAS.487.4136W}.  For eRO-QPE1 and eRO-QPE2, since the
masses of their host galaxies have been derived in previous works
\citep{2021Natur.592..704A}, we use them to estimate the masses of the MBHs
according to the empirical scaling relation
\begin{equation}
\log(M_{\mathrm{BH}}/M_{\odot}) = 7.45 + 1.05\log(M_{\mathrm{stellar}}/10^{11}M_{\odot})
\end{equation}
\citep{reines15}.

The mass $m$ of the small object is model-dependent and uncertain. To
accommodate various theoretical possibilities, we treat $m$ as a free parameter and vary
it between $0.2 M_\odot$, mimicing WDs or stripped cores of main-sequence
stars, to $10 M_\odot$, accounting for massive main-sequence stars or
stellar-mass BHs. 

Given the above parameters and assume that the orbits are circular, the GW
radiation timescale \citep{1964PhRv..136.1224P} is many orders of magnitude
longer than the mission duration of LISA, about $t_{\rm LISA}=4$ years. In this
case, the GW spectrum is essentially monochromatic and the increment of
frequency $\Delta f$ during the observational period  is much smaller than the
GW frequency $f$. Note that for circular orbits $f=2/P$.

In this situation, the SNR can be calculated with
\begin{equation}
	{\rm SNR}^2=\frac{h_c^2\Delta f}{f^2S(f)},\label{eq:SNR}
\end{equation}
where $h_c$ is the characteristic strain and $S(f)$ is the one-side
amplitude spectral density of LISA \citep{robson19}. 
According to the last equation, the effective strain, which is directly
proportional to the SNR, is 
\begin{equation}
	h_{\rm eff}=h_c\sqrt{\Delta f/f}
\end{equation}
\citep[also see][]{barack04}.
It is smaller than the normally adopted characteristic strain $h_c$ of a fast
chirping signal by a factor of 
$\sqrt{\Delta f/f}$.

\begin{figure}
\centering
\includegraphics[width=0.5\textwidth]{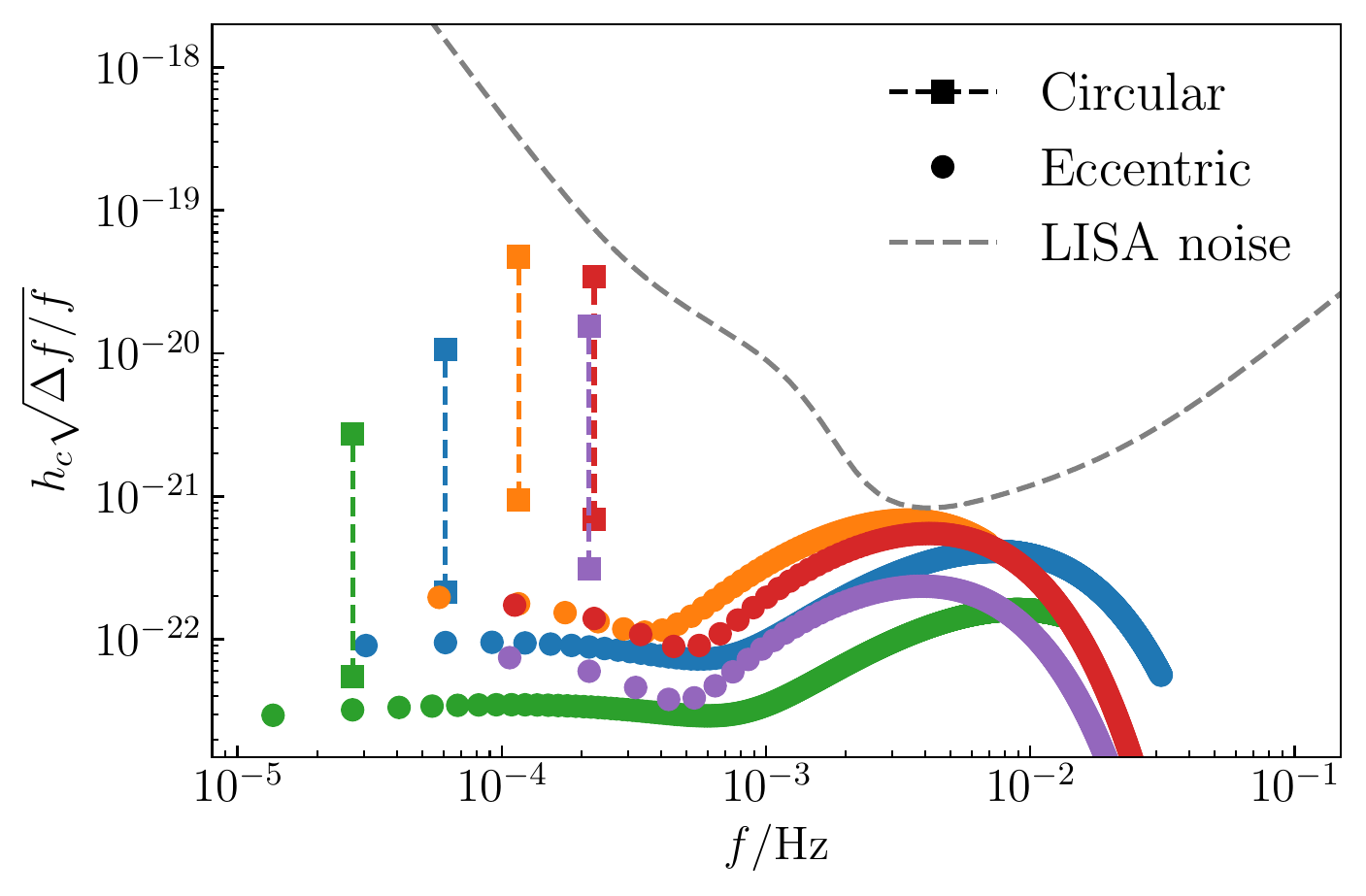}
	\caption{The effective GW strain of QPEs (colored squares and solid dots) versus the LISA sensitivity curve $\sqrt{fS(f)}$ (grey dashed line). 
The squares connected by dashed lines correspond to the model in which QPEs contain
circular binaries. The dots refer to the model in which QPEs are
powered by WDs on eccentric orbits around MBHs.
\label{fig:models}}
\end{figure}

The effective strain computed using the above model and assuming $t_{\rm
LISA}=4$ years is shown in Figure~\ref{fig:models} as the squares connected by
dashed lines.  Each QPE is represented by a vertical line segment because we
have allowed $m$ to vary between $0.2$ and $10 M_\odot$. They are all below the
LISA sensitively curve $\sqrt{fS(f)}$, suggesting that LISA could not detect
such QPEs. 

\subsection{WDs on eccentric orbits}
\label{sec:ecc}

We now consider another model in which
the small object in a QPE is a WD and it is moving along
a highly eccentric orbit around the central MBH
\citep{2020MNRAS.493L.120K}. 
In this model, GW radiation causes the orbital
pericenter to decay until the WD fills the Roche lobe and
starts feeding the MBH.
Such a system can be characterized by four parameters:
besides $M$ of the MBH and $m$ of the WD, there are also the semimajor axis
$a$ and eccentricity $e$ of the orbit.
Following \citet{2020MNRAS.493L.120K}, we derive these parameters
using their relationships with the observables of QPEs.
The basic steps are given below.

We first adopt the MBH mass $M$ and orbital period $P$ from the previous
subsection.  Then the semimajor axis can be derived from
$P=2\pi(GM/a^3)^{-1/2}$.  To establish a relationship between the remaining two
parameters, $m$ and $e$, we use the physical requirement that the mass-transfer
timescale $m/\dot{M}$ equals the decay timescale of the pericenter
$|r_p/\dot{r}_p|$ , where $\dot{M}$ is the orbit-averaged accretion rate of the
MBH and $r_p=a(1-e)$ is the pericenter distance.  Throughout this paper, the
dot symbol denotes the time derivative. 

The accretion rate $\dot{M}$ is determined by the light curve of the eruptions.
From the peak luminosity $L$ of an eruption and its full width at half-maximum $\Delta t$, 
we get $\dot{M}=L\Delta
t/(\eta Pc^2)$, where $\eta$ is the radiative efficiency and $c$ is the speed
of light.  To write $|r_p/\dot{r}_p|$ in terms of $a$ and $e$, we note that the
specific angular momentum $J=\sqrt{G(M+m)a(1-e^2)}$ is proportional to
$\sqrt{r_p}$ when $e\simeq1$.  Therefore, we can write $m/\dot{M}\simeq|J/\dot{J}|$.
We note that according to \citet{1964PhRv..136.1224P}, $|J/\dot{J}|$ is
proportional to $(1-e^2)^{5/2}$, not $(1-e^2)^{7/2}$ which is related to the
loss of orbital energy and has been misused in the previous works
\citep[e.g.][]{2020MNRAS.493L.120K,2021arXiv210903471Z}.

To close the equations we need another relationship between $m$ and $e$.
This is given by the condition of Roche-lobe overflow. It requires that 
during the pericenter passage
the WD, which has a size of about $0.013R_\odot(m/M_\odot)^{-1/3}$,
fills the Roche lobe, whose radius is $0.46r_p(m/M)^{1/3}$ \citep{2020MNRAS.493L.120K}.
Finally, we find that
\begin{align}
	m&\simeq0.20C^{-15/22}M_\odot,\\
	e&\simeq1-0.072C^{5/11}P_4^{-2/3},
\end{align}
where
\begin{equation}
	C=\left(\frac{M}{10^5M_\odot}\right)^{4/15}
	\left(\frac{L\,\Delta t}{10^{45}{\rm erg}}\right)^{-2/5}
	\left(\frac{\eta}{0.1}\right)^{2/5}
\end{equation}
and $P_4=P/(10^4\,{\rm s})$.

The values of the observables, $\Delta t$ and $L$, as well as the derived
physical parameters, $m$ and $e$, are given in Table~\ref{tab:sample}. We
find that $m$ falls in the typical mass range of WDs, suggesting that the
model is self-consistent.  Moreover, $e$ is higher than $0.9$, consistent with
the scenario that the WDs are delivered to the MBHs by either partial tidal
disruption or binary separation, though the event rate is expected to be low
\citep{2021arXiv210713015M}.

Because the binary is now eccentric, the GW radiation is spread into a wide
range of harmonic frequencies \citep{peters63}.  We use the formulae derived in
\citet{barack04} to compute the characteristic strain $h_{c,n}$ of the $n$th
harmonic, which is at the frequency $n/P$.  The effective strain, which is
directly correlated with the SNR, is computed with $h_{c,n}\sqrt{\Delta f/f}$
\citep{barack04} and shown in Figure~\ref{fig:models} as the colored dots.

It is clear that each QPE is now emitting a wide GW spectrum.  We find that the
peak of the spectrum occurs at a frequency of about $\sqrt{GM/r_p^3}$, which
can be understood due to the fact that the strongest GW radiation is produced
when the WD passes the orbital pericenter.  We also find that the frequency of
the peak coincides with the most sensitive band of LISA, around $3$ milli-Hertz (mHz).
However, the effective strain remains below the sensitivity curve of LISA,
indicating that LISA cannot detect an individual QPE with sufficient SNR.

\section{GW background}\label{sec:GWbg}

Although an individual QPE is too weak to detect by LISA, several QPEs together
may increase the SNR.  This could happen if QPEs are made of WDs on eccentric
orbits around MBHs.  In this case, the GW spectrum is broad, as we have seen in
the previous section.  The broadness increases the chance of signal overlapping
at the same frequency.  For this reason, we use the model described in
Section~\ref{sec:ecc} to estimate the combined GW signal of many QPEs.

First, we determine how many QPEs exist per unit comoving volume.  Based on the
five QPEs detected so far (Table~\ref{tab:sample}), we infer a comoving number
density of $n_*\simeq120\,{\rm Gpc}^{-3}$.  We use this value for the later
calculation of the GW background. However, we caution that it should be
regarded as a lower limit for two reasons. (i) The current QPE sample is by no
means complete because it is compiled from heterogeneous observations.  (ii)
Many more WD-MBH binaries are expected to reside on wider orbits because the
evolution timescale (GW radiation timescale) is longer for wider binaries.
Wide binaries also emit GWs but not necessary X-rays because 
the mass transfer may not have started. Therefore, they are not included
in the QPE sample. 

Second, assuming that the comoving number density $n$ does not evolve with
redshift, we generate a mock sample of $18,000$ QPEs in the redshift range of
$0\le z\le1$.  We do not consider higher redshift mainly because our result
converges as $z$ approaches $1$, as we will see later.  For each mock QPE, we
randomly choose one of the five detected QPEs in Table~\ref{tab:sample} and
assign the parameters $M,m,P,e$ of the selected QPE to the mock one.  Having
specified the parameters, we then compute the effective strain 
as is seen by LISA \citep[following][]{barack04}.  In the calculation, we
assume $t_{\rm LISA}=4$ years. 
   
\begin{figure}
\centering
\includegraphics[width=0.5\textwidth]{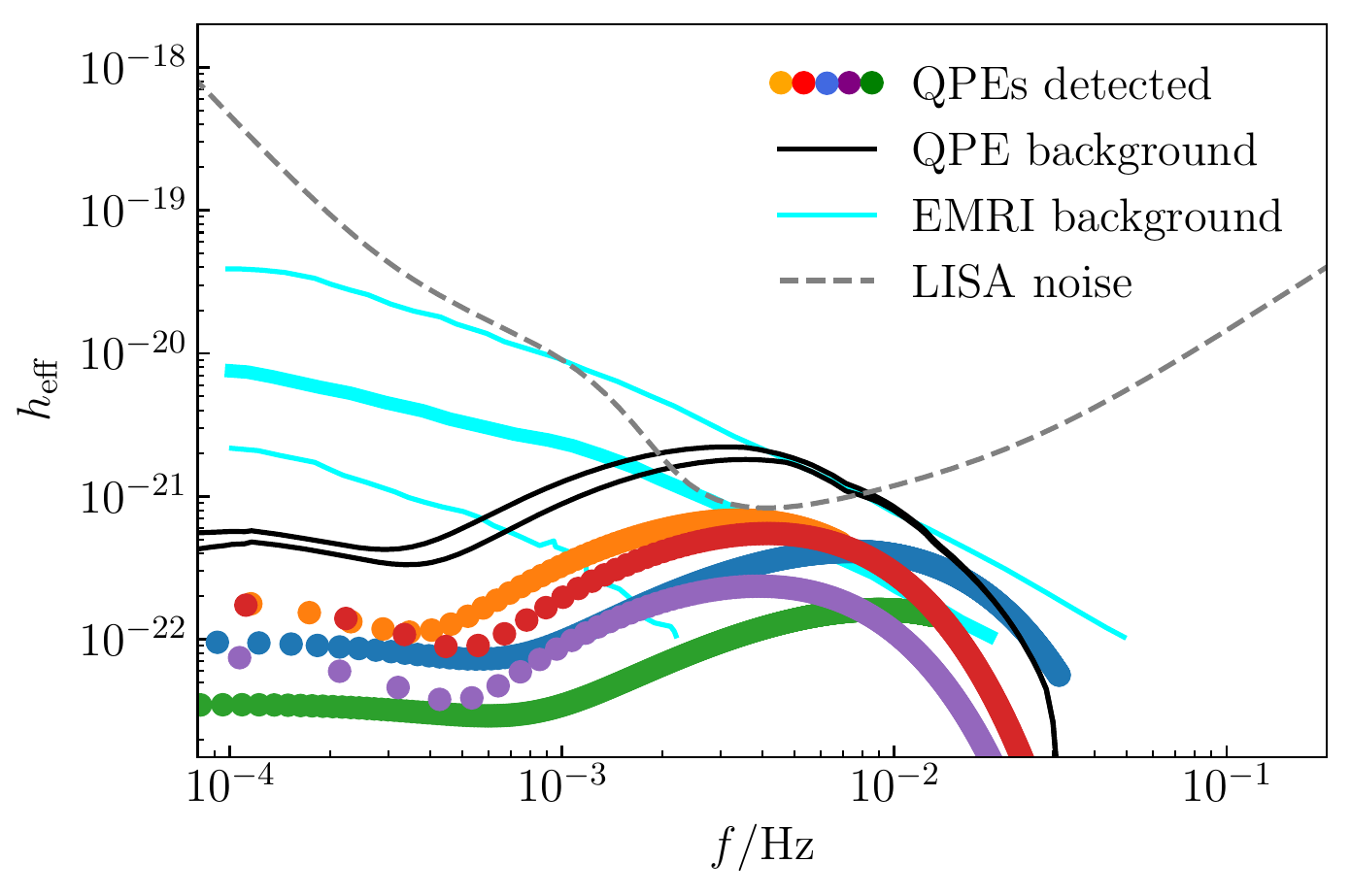}
\caption{The minimum GW background produced by QPEs (black solid curves).
There are two black curves because they refer to the contribution from the QPEs at, respectively,
$z\le0.5$ and $z\le1$. 
The effective strain of the five detected
QPEs are also shown (colored dots) for comparison.
The three cyan curves show the background produced by
the EMRIs containing stellar-mass BHs, and from top to bottom
they refer to
the most optimistic, fiducial, and the most pessimistic
estimations \citep[from][]{bonetti20}. 
\label{fig:QPEbg}}
\end{figure}

Third, the GW signals of different QPEs add up incoherently, so the total
effective strain $h_{\rm eff}$ is computed with
\begin{equation} h^2_{\rm eff}(f)=\sum_{i} h^2_{{\rm eff},i}(f), \end{equation}
where $h_{{\rm eff},i}$ denotes the effective strain of the $i$th QPE of our
mock sample.  The result is shown in Figure~\ref{fig:QPEbg} as the two black solid
curves, which refer to the background produced by the QPEs at, respectively,
$z\le0.5$ and $z\le1$.  The corresponding SNR is $2.1$ and $2.5$, where we
have calculated the total SNR of the GW background with  
\begin{equation} {\rm SNR}^2=\int\frac{h_{\rm eff}^2(f)
}{fS(f)}d\ln f. \end{equation}

We emphasize that the QPE background derived here should be regarded as a lower
limit because it is computed based on the most conservative estimation of the
number density of QPEs, i.e., only five QPEs within a redshift of $0.05$.
Nevertheless, we find that in the frequency band of $4-20$ mHz the QPE
background is comparable to the most optimistic estimation of the GW background
produced by those EMRIs containing stellar-mass BHs (see the highest cyan curve
in Fig.~\ref{fig:QPEbg}). The QPE background is also orders of magnitude higher
than the GW background generated by tidal disruption events \citep[see][not
shown here]{toscani20}. Therefore, we conclude that QPE is an important source
of stochastic GW background in the mHz band, in fact the most sensitive band of
LISA.

\begin{figure}
\centering
\includegraphics[width=0.5\textwidth]{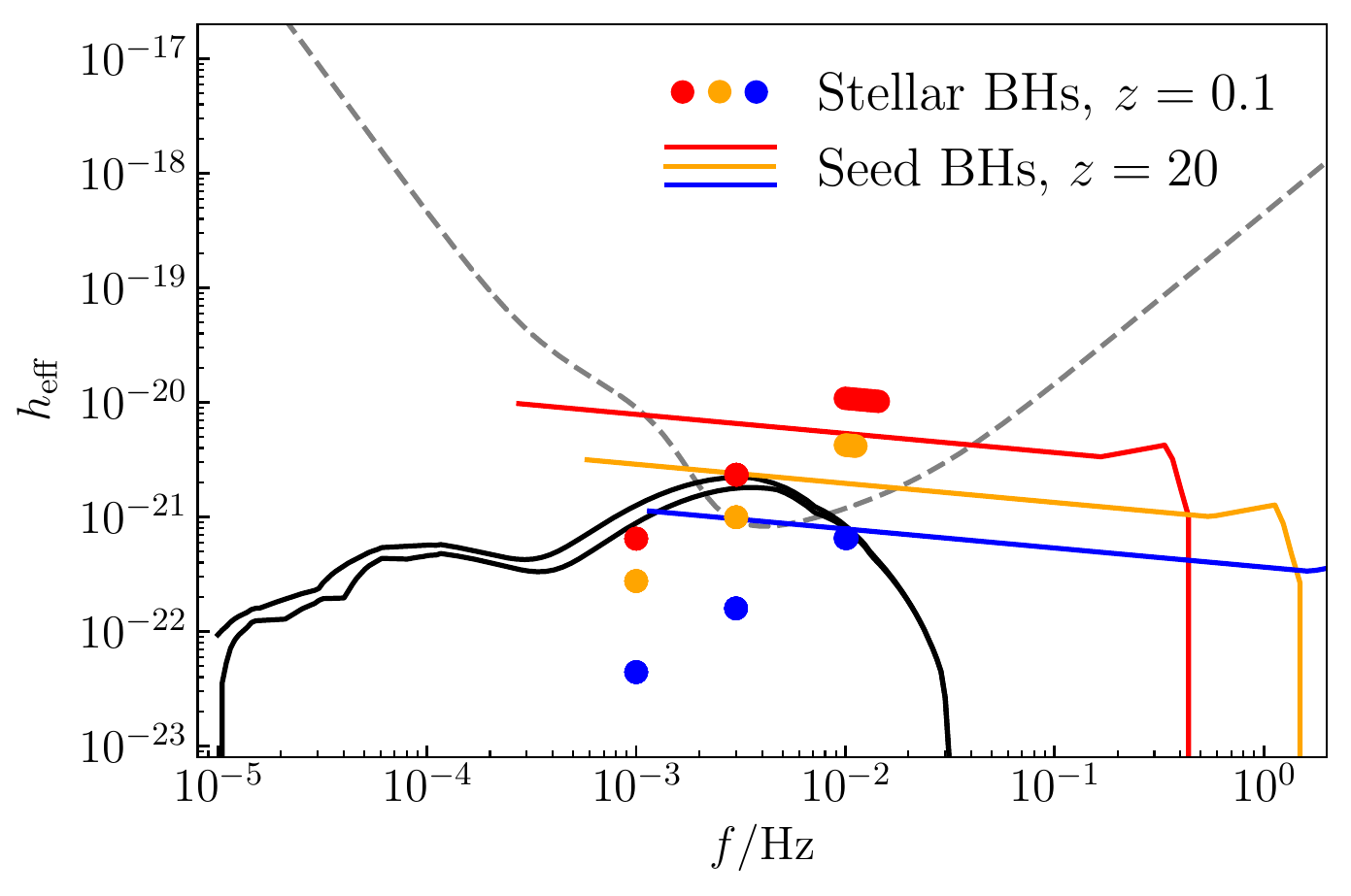}
\caption{Comparing the GW background produced by QPEs (black solid curves) 
with other GW sources. The colored curves represent the
chirping signals of merging seed BHs at redshift $z=20$. 
From top to bottom, the red, orange, and blue curves correspond to
a chirp mass of, respectively, $1000$, $300$, and $100\,M_\odot$.
The colored dots show the effective strain of inspiraling stellar-mass binary BHs
residing at three representative frequencies
$(1, 3, 10)$ mHz
and a redshift of $z=0.1$. 
The red, orange, and blue dots correspond to a chirp mass of 
$50$, $30$, and $10\,M_\odot$.
An observational period of 
$t_{\rm LISA}=4$ years is assumed in the calculation.
	\label{fig:comp}}
\end{figure}

\citet{bonetti20} pointed out that an excessive GW background in the mHz GW
band would impinge on several science goals of LISA, including the search for
seed MBHs at $z\ga20$ as well as detecting stellar-mass BBHs in their
early inspiral phase. To
understand the ramification of our results, we show in Figure~\ref{fig:comp}
the GW signals of seed MBHs and stellar-mass BBHs and compare them with the
background due to QPEs. We find that the QPE background is higher than the
chirp signal of a MBH binary at $z=20$, if the chirp mass is lower than $300
M_\odot$. Such a mass corresponds to the seed BHs produced by population-III
stars \citep{volonteri10}.  We also find that at $f\la3$ mHz the QPE background
becomes higher than the effective strain of a stellar-mass BBH at $z=0.1$, if
the chirp mass is smaller than $50 M_\odot$. The progenitors of many LIGO/Virgo
BBHs fall in this mass range. These results highlight the necessity of
observationally compiling a complete sample of QPEs to put a better constraint
on the level of the GW background.

\section{Summary and Discussion}\label{sec:dis}

Motivated by the suggestion that the recently discovered QPEs may contain
EMRIs, we have calculated in this paper the GW spectra and studied their
detectability by LISA.  We investigated two scenarios proposed in the
literature, in which the orbits of the EMRIs are, respectively, circular and
highly eccentric ($e>0.9$).  We found that in both cases the signal of an
individual QPE is too weak to be discernible by LISA (Fig.~\ref{fig:models}).

This conclusion differs from the one made by \citet{2021arXiv210903471Z},
mainly because we noticed that the systems are evolving on a timescale much
longer than the mission duration of LISA, so that the number of GW cycles
accumulated in the LISA band is greatly suppressed relative to the number of
cycles coming from fast-chirping sources.  
\citet{sesana08,han18} also studied the GW signal of a WD around
an intermediate-massive BH (IMBH). We notice that the SNR derived by them is
much higher than the detection threshold of LISA.  This result is caused by the
fact that their WDs are only a few years away from the final merger with the
IMBHs, while our QPEs are thousands of years before the merger. The GW
radiation is much weaker in our case. 

More importantly, we found that if the EMRIs in QPEs are eccentric, their broad GW
spectra could overlap in a wide range of frequencies, producing a background
which has a much higher SNR.  We showed that
this background is already higher than the LISA noise curve even if we adopt the
most conservative assumption about the abundance of QPEs (Fig.~\ref{fig:QPEbg}).
Moreover, in the frequency band of $4-20$ mHz, the QPE minimum background is
comparable to the most optimistic estimation of the background from the fiducial EMRIs, i.e.,
those EMRIs containing stellar-mass BHs. This result implies that QPE may be
the dominant source of confusion noise in the most sensitive band of LISA.
We have shown that its presence may affect the future search for 
seed BHs and stellar-mass BBHs by LISA (Fig.~\ref{fig:comp}).

When calculating the GW background, we mentioned that our model missed the
WD-MBH binaries on wide orbits (because they have not started mass transfer and
will not be detected as QPEs). We can roughly estimate the contribution of
these binaries to the SNR of the GW background.  Suppose that within the
orbital period of $P$ to $P+dP$ there are a number of $dN$ WD-MBH binaries.
The number density, $dN/dP$, is proportional to $1/|\dot{P}|$ according to the
continuity equation.  This relation leads to $dN/d\ln P\propto |P/\dot{P}|$.
In our problem, $\dot{P}$ is determined by GW radiation, and during the
evolution of $P$, the pericenter distance $r_p$ is more or less conserved, due to the
fact that $|P/\dot{P}|/|r_p/\dot{r}_p|\sim (1-e)\ll1$. In this case, we can
derive $dN/d\ln P\propto r_p^{7/2}P^{1/3}$, where we have used the relation
$P/\dot{P}\propto a/\dot{a}\propto a^4(1-e)^{7/2}$ \citep{1964PhRv..136.1224P}.
If we can further estimate the effective strain $h_{\rm eff}$ of such a WD-MBH
binary, we can derive the SNR contributed by these $dN/d\ln P$ binaries as
$d({\rm SNR}^2)/d\ln P\sim h^2_{\rm eff} (dN/d\ln P)$.  For $h_{\rm eff}$, we
have seen that it peaks at a frequency of $f_p\sim(GM/r_p^3)^{1/2}$,
due to the fact that most of the GW energy is radiated during the 
pericenter passage. Therefore, the corresponding
effective strain can be calculated with $h^2_{\rm eff}\propto (\dot{E}_{\rm
tot}/\dot{f}_p)(\dot{f}_p/f_p)t_{\rm LISA}$, where $\dot{E}_{\rm tot}$ is the
total power of GW radiation \citep{barack04}. Finally, we find that $h^2_{\rm
eff}\propto P^{-1}r_p^{-2}t_{\rm LISA}$, and $d({\rm SNR}^2)/d\ln P\sim
P^{-2/3}r_p^{3/2}t_{\rm LISA}$. The last equation indicates that the binary
population with longer orbital period contribute less SNR to the GW
background.  Therefore, neglecting wide WD-MBHs should not qualitatively
change the result about the GW background.

Finally, we point out two caveats of this work.
First, the GW background derived in this work should be regarded as a lower
limit because the calculation is based on the five QPEs detected so far.  If
more QPEs would be discovered in the future, the GW background would increase
as $\sqrt{n_*}$. Second, whether QPEs are made of WDs moving on eccentric
orbits around MBHs is still unclear. Further theoretical work is needed to 
identify observable signatures that can be used to distinguish different
models.

\acknowledgments

This work is supported by the National Science Foundation of China grants No 11991053 and 11873022, and the China Manned Spaced Project (CMS-CSST-2021-B11).

\bibliographystyle{aasjournal1}
\bibliography{ref}

\begin{thebibliography}{}
\expandafter\ifx\csname natexlab\endcsname\relax\def\natexlab#1{#1}\fi
\providecommand{\url}[1]{\href{#1}{#1}}
\providecommand{\dodoi}[1]{doi:~\href{http://doi.org/#1}{\nolinkurl{#1}}}
\providecommand{\doeprint}[1]{\href{http://ascl.net/#1}{\nolinkurl{http://ascl.net/#1}}}
\providecommand{\doarXiv}[1]{\href{https://arxiv.org/abs/#1}{\nolinkurl{https://arxiv.org/abs/#1}}}

\bibitem[{{Amaro-Seoane} {et~al.}(2007){Amaro-Seoane}, {Gair}, {Freitag},
  {Miller}, {Mandel}, {Cutler}, \& {Babak}}]{amaro-seoane07}
{Amaro-Seoane}, P., {Gair}, J.~R., {Freitag}, M., {et~al.} 2007, Classical and
  Quantum Gravity, 24, R113, \dodoi{10.1088/0264-9381/24/17/R01}

\bibitem[{{Amaro-Seoane} {et~al.}(2017){Amaro-Seoane}, {Audley}, {Babak},
  {Baker}, {Barausse}, {Bender}, {Berti}, {Binetruy}, {Born}, {Bortoluzzi},
  {Camp}, {Caprini}, {Cardoso}, {Colpi}, {Conklin}, {Cornish}, {Cutler},
  {Danzmann}, {Dolesi}, {Ferraioli}, {Ferroni}, {Fitzsimons}, {Gair}, {Gesa
  Bote}, {Giardini}, {Gibert}, {Grimani}, {Halloin}, {Heinzel}, {Hertog},
  {Hewitson}, {Holley-Bockelmann}, {Hollington}, {Hueller}, {Inchauspe},
  {Jetzer}, {Karnesis}, {Killow}, {Klein}, {Klipstein}, {Korsakova}, {Larson},
  {Livas}, {Lloro}, {Man}, {Mance}, {Martino}, {Mateos}, {McKenzie},
  {McWilliams}, {Miller}, {Mueller}, {Nardini}, {Nelemans}, {Nofrarias},
  {Petiteau}, {Pivato}, {Plagnol}, {Porter}, {Reiche}, {Robertson},
  {Robertson}, {Rossi}, {Russano}, {Schutz}, {Sesana}, {Shoemaker}, {Slutsky},
  {Sopuerta}, {Sumner}, {Tamanini}, {Thorpe}, {Troebs}, {Vallisneri},
  {Vecchio}, {Vetrugno}, {Vitale}, {Volonteri}, {Wanner}, {Ward}, {Wass},
  {Weber}, {Ziemer}, \& {Zweifel}}]{lisa17}
{Amaro-Seoane}, P., {Audley}, H., {Babak}, S., {et~al.} 2017, arXiv e-prints,
  arXiv:1702.00786.
\newblock \doarXiv{1702.00786}

\bibitem[{{Arcodia} {et~al.}(2021){Arcodia}, {Merloni}, {Nandra}, {Buchner},
  {Salvato}, {Pasham}, {Remillard}, {Comparat}, {Lamer}, {Ponti}, {Malyali},
  {Wolf}, {Arzoumanian}, {Bogensberger}, {Buckley}, {Gendreau}, {Gromadzki},
  {Kara}, {Krumpe}, {Markwardt}, {Ramos-Ceja}, {Rau}, {Schramm}, \&
  {Schwope}}]{2021Natur.592..704A}
{Arcodia}, R., {Merloni}, A., {Nandra}, K., {et~al.} 2021, \nat, 592, 704,
  \dodoi{10.1038/s41586-021-03394-6}

\bibitem[{{Babak} {et~al.}(2017){Babak}, {Gair}, {Sesana}, {Barausse},
  {Sopuerta}, {Berry}, {Berti}, {Amaro-Seoane}, {Petiteau}, \&
  {Klein}}]{babak17}
{Babak}, S., {Gair}, J., {Sesana}, A., {et~al.} 2017, \prd, 95, 103012,
  \dodoi{10.1103/PhysRevD.95.103012}

\bibitem[{{Barack} \& {Cutler}(2004)}]{barack04}
{Barack}, L., \& {Cutler}, C. 2004, \prd, 70, 122002,
  \dodoi{10.1103/PhysRevD.70.122002}

\bibitem[{{Berry} {et~al.}(2019){Berry}, {Hughes}, {Sopuerta}, {Chua},
  {Heffernan}, {Holley-Bockelmann}, {Mihaylov}, {Miller}, \&
  {Sesana}}]{berry19}
{Berry}, C., {Hughes}, S., {Sopuerta}, C., {et~al.} 2019, \baas, 51, 42.
\newblock \doarXiv{1903.03686}

\bibitem[{{Bogdanovi{\'c}} {et~al.}(2014){Bogdanovi{\'c}}, {Cheng}, \&
  {Amaro-Seoane}}]{bogdanovic14}
{Bogdanovi{\'c}}, T., {Cheng}, R.~M., \& {Amaro-Seoane}, P. 2014, \apj, 788,
  99, \dodoi{10.1088/0004-637X/788/2/99}

\bibitem[{{Bonetti} \& {Sesana}(2020)}]{bonetti20}
{Bonetti}, M., \& {Sesana}, A. 2020, \prd, 102, 103023,
  \dodoi{10.1103/PhysRevD.102.103023}

\bibitem[{{Chakraborty} {et~al.}(2021){Chakraborty}, {Kara}, {Masterson},
  {Giustini}, {Miniutti}, \& {Saxton}}]{2021ApJ...921L..40C}
{Chakraborty}, J., {Kara}, E., {Masterson}, M., {et~al.} 2021, \apjl, 921, L40,
  \dodoi{10.3847/2041-8213/ac313b}

\bibitem[{{de Freitas Pacheco} {et~al.}(2006){de Freitas Pacheco}, {Filloux},
  \& {Regimbau}}]{pacheco06}
{de Freitas Pacheco}, J.~A., {Filloux}, C., \& {Regimbau}, T. 2006, \prd, 74,
  023001, \dodoi{10.1103/PhysRevD.74.023001}

\bibitem[{{Gair} {et~al.}(2017){Gair}, {Babak}, {Sesana}, {Amaro-Seoane},
  {Barausse}, {Berry}, {Berti}, \& {Sopuerta}}]{gair17}
{Gair}, J.~R., {Babak}, S., {Sesana}, A., {et~al.} 2017, in Journal of Physics
  Conference Series, Vol. 840, Journal of Physics Conference Series, 012021,
  \dodoi{10.1088/1742-6596/840/1/012021}

\bibitem[{{Gair} {et~al.}(2013){Gair}, {Vallisneri}, {Larson}, \&
  {Baker}}]{gair13}
{Gair}, J.~R., {Vallisneri}, M., {Larson}, S.~L., \& {Baker}, J.~G. 2013,
  Living Reviews in Relativity, 16, 7, \dodoi{10.12942/lrr-2013-7}

\bibitem[{{Giustini} {et~al.}(2020){Giustini}, {Miniutti}, \&
  {Saxton}}]{giustini20}
{Giustini}, M., {Miniutti}, G., \& {Saxton}, R.~D. 2020, \aap, 636, L2,
  \dodoi{10.1051/0004-6361/202037610}

\bibitem[{{Han} \& {Fan}(2018)}]{han18}
{Han}, W.-B., \& {Fan}, X.-L. 2018, \apj, 856, 82,
  \dodoi{10.3847/1538-4357/aab03c}

\bibitem[{{Hils} \& {Bender}(1995)}]{hils95}
{Hils}, D., \& {Bender}, P.~L. 1995, \apjl, 445, L7, \dodoi{10.1086/187876}

\bibitem[{{Ingram} {et~al.}(2021){Ingram}, {Motta}, {Aigrain}, \&
  {Karastergiou}}]{2021MNRAS.503.1703I}
{Ingram}, A., {Motta}, S.~E., {Aigrain}, S., \& {Karastergiou}, A. 2021,
  \mnras, 503, 1703, \dodoi{10.1093/mnras/stab609}

\bibitem[{{Ivanov}(2002)}]{ivanov02}
{Ivanov}, P.~B. 2002, \mnras, 336, 373,
  \dodoi{10.1046/j.1365-8711.2002.05733.x}

\bibitem[{{Ivanov} \& {Papaloizou}(2007)}]{ivanov07}
{Ivanov}, P.~B., \& {Papaloizou}, J.~C.~B. 2007, \aap, 476, 121,
  \dodoi{10.1051/0004-6361:20077105}

\bibitem[{{King}(2020)}]{2020MNRAS.493L.120K}
{King}, A. 2020, \mnras, 493, L120, \dodoi{10.1093/mnrasl/slaa020}

\bibitem[{{Luminet} \& {Pichon}(1989)}]{luminet89}
{Luminet}, J.~P., \& {Pichon}, B. 1989, \aap, 209, 103

\bibitem[{{MacLeod} {et~al.}(2014){MacLeod}, {Goldstein}, {Ramirez-Ruiz},
  {Guillochon}, \& {Samsing}}]{macleod14}
{MacLeod}, M., {Goldstein}, J., {Ramirez-Ruiz}, E., {Guillochon}, J., \&
  {Samsing}, J. 2014, \apj, 794, 9, \dodoi{10.1088/0004-637X/794/1/9}

\bibitem[{{Metzger} {et~al.}(2021){Metzger}, {Stone}, \&
  {Gilbaum}}]{2021arXiv210713015M}
{Metzger}, B.~D., {Stone}, N.~C., \& {Gilbaum}, S. 2021, arXiv e-prints,
  arXiv:2107.13015.
\newblock \doarXiv{2107.13015}

\bibitem[{{Miller} {et~al.}(2005){Miller}, {Freitag}, {Hamilton}, \&
  {Lauburg}}]{miller05}
{Miller}, M.~C., {Freitag}, M., {Hamilton}, D.~P., \& {Lauburg}, V.~M. 2005,
  \apjl, 631, L117, \dodoi{10.1086/497335}

\bibitem[{{Miniutti} {et~al.}(2019){Miniutti}, {Saxton}, {Giustini},
  {Alexander}, {Fender}, {Heywood}, {Monageng}, {Coriat}, {Tzioumis}, {Read},
  {Knigge}, {Gandhi}, {Pretorius}, \&
  {Ag{\'\i}s-Gonz{\'a}lez}}]{2019Natur.573..381M}
{Miniutti}, G., {Saxton}, R.~D., {Giustini}, M., {et~al.} 2019, \nat, 573, 381,
  \dodoi{10.1038/s41586-019-1556-x}

\bibitem[{{Motta} {et~al.}(2020){Motta}, {Marelli}, {Pintore}, {Esposito},
  {Salvaterra}, {De Luca}, {Israel}, {Tiengo}, \& {Castillo}}]{motta20}
{Motta}, S.~E., {Marelli}, M., {Pintore}, F., {et~al.} 2020, \apj, 898, 174,
  \dodoi{10.3847/1538-4357/ab9b81}

\bibitem[{{Peters}(1964)}]{1964PhRv..136.1224P}
{Peters}, P.~C. 1964, Physical Review, 136, 1224,
  \dodoi{10.1103/PhysRev.136.B1224}

\bibitem[{{Peters} \& {Mathews}(1963)}]{peters63}
{Peters}, P.~C., \& {Mathews}, J. 1963, Physical Review, 131, 435,
  \dodoi{10.1103/PhysRev.131.435}

\bibitem[{{Raj} \& {Nixon}(2021)}]{2021ApJ...909...82R}
{Raj}, A., \& {Nixon}, C.~J. 2021, \apj, 909, 82,
  \dodoi{10.3847/1538-4357/abdc25}

\bibitem[{{Reines} \& {Volonteri}(2015)}]{reines15}
{Reines}, A.~E., \& {Volonteri}, M. 2015, \apj, 813, 82,
  \dodoi{10.1088/0004-637X/813/2/82}

\bibitem[{{Robson} {et~al.}(2019){Robson}, {Cornish}, \& {Liu}}]{robson19}
{Robson}, T., {Cornish}, N.~J., \& {Liu}, C. 2019, Classical and Quantum
  Gravity, 36, 105011, \dodoi{10.1088/1361-6382/ab1101}

\bibitem[{{Rosswog} {et~al.}(2009){Rosswog}, {Ramirez-Ruiz}, \&
  {Hix}}]{rosswog09}
{Rosswog}, S., {Ramirez-Ruiz}, E., \& {Hix}, W.~R. 2009, \apj, 695, 404,
  \dodoi{10.1088/0004-637X/695/1/404}

\bibitem[{{Sesana} {et~al.}(2008){Sesana}, {Vecchio}, {Eracleous}, \&
  {Sigurdsson}}]{sesana08}
{Sesana}, A., {Vecchio}, A., {Eracleous}, M., \& {Sigurdsson}, S. 2008, \mnras,
  391, 718, \dodoi{10.1111/j.1365-2966.2008.13904.x}

\bibitem[{{Sheng} {et~al.}(2021){Sheng}, {Wang}, {Ferland}, {Shu}, {Yang},
  {Jiang}, \& {Chen}}]{sheng21}
{Sheng}, Z., {Wang}, T., {Ferland}, G., {et~al.} 2021, \apjl, 920, L25,
  \dodoi{10.3847/2041-8213/ac2251}

\bibitem[{{Shu} {et~al.}(2017){Shu}, {Wang}, {Jiang}, {Wang}, {Sun}, \&
  {Zhou}}]{2017ApJ...837....3S}
{Shu}, X.~W., {Wang}, T.~G., {Jiang}, N., {et~al.} 2017, \apj, 837, 3,
  \dodoi{10.3847/1538-4357/aa5eb3}

\bibitem[{{Sigl} {et~al.}(2007){Sigl}, {Schnittman}, \& {Buonanno}}]{sigl07}
{Sigl}, G., {Schnittman}, J., \& {Buonanno}, A. 2007, \prd, 75, 024034,
  \dodoi{10.1103/PhysRevD.75.024034}

\bibitem[{{Sigurdsson} \& {Rees}(1997)}]{sigurdsson97}
{Sigurdsson}, S., \& {Rees}, M.~J. 1997, \mnras, 284, 318,
  \dodoi{10.1093/mnras/284.2.318}

\bibitem[{{Sniegowska} {et~al.}(2020){Sniegowska}, {Czerny}, {Bon}, \&
  {Bon}}]{sniegowska20}
{Sniegowska}, M., {Czerny}, B., {Bon}, E., \& {Bon}, N. 2020, \aap, 641, A167,
  \dodoi{10.1051/0004-6361/202038575}

\bibitem[{{Toscani} {et~al.}(2020){Toscani}, {Rossi}, \& {Lodato}}]{toscani20}
{Toscani}, M., {Rossi}, E.~M., \& {Lodato}, G. 2020, \mnras, 498, 507,
  \dodoi{10.1093/mnras/staa2290}

\bibitem[{{Volonteri}(2010)}]{volonteri10}
{Volonteri}, M. 2010, \aapr, 18, 279, \dodoi{10.1007/s00159-010-0029-x}

\bibitem[{{Wevers} {et~al.}(2019){Wevers}, {Stone}, {van Velzen}, {Jonker},
  {Hung}, {Auchettl}, {Gezari}, {Onori}, {Mata S{\'a}nchez},
  {Kostrzewa-Rutkowska}, \& {Casares}}]{2019MNRAS.487.4136W}
{Wevers}, T., {Stone}, N.~C., {van Velzen}, S., {et~al.} 2019, \mnras, 487,
  4136, \dodoi{10.1093/mnras/stz1602}

\bibitem[{{Xian} {et~al.}(2021){Xian}, {Zhang}, {Dou}, {He}, \&
  {Shu}}]{2021ApJ...921L..32X}
{Xian}, J., {Zhang}, F., {Dou}, L., {He}, J., \& {Shu}, X. 2021, \apjl, 921,
  L32, \dodoi{10.3847/2041-8213/ac31aa}

\bibitem[{{Zalamea} {et~al.}(2010){Zalamea}, {Menou}, \&
  {Beloborodov}}]{zalamea10}
{Zalamea}, I., {Menou}, K., \& {Beloborodov}, A.~M. 2010, \mnras, 409, L25,
  \dodoi{10.1111/j.1745-3933.2010.00930.x}

\bibitem[{{Zhao} {et~al.}(2021){Zhao}, {Wang}, {Zou}, {Wang}, \&
  {Dai}}]{2021arXiv210903471Z}
{Zhao}, Z.~Y., {Wang}, Y.~Y., {Zou}, Y.~C., {Wang}, F.~Y., \& {Dai}, Z.~G.
  2021, arXiv e-prints, arXiv:2109.03471.
\newblock \doarXiv{2109.03471}

\end{thebibliography}

\end{document}